\documentstyle{aipproc}

\begin{document}
\title{Lorentz Symmetry Violation and Acceleration in Relativistic Shocks}

\author{Luis Gonzalez-Mestres}
\address{CNRS-IN2P3, B.P. 110, 74941 Annecy-le-Vieux Cedex, France \\ 
E-mail: gonzalez@lappa.in2p3.fr , telephone and fax: 33 1 45830720}
\maketitle

\begin{abstract}
In previous papers (e.g. physics/0003080 and references therein), it 
was pointed out that Lorentz symmetry violation (LSV) at
Planck scale can play a crucial role in astrophysical processes at very
high energy. It can potentially
inhibit radiation under
external forces (e.g. synchrotron-like), GZK-like cutoffs, decays,
photodisintegration of nuclei, momentum loss trough
collisions (e.g. with a photon wind in reverse shocks),
production of lower-energy secondaries...
An updated description of this phenomenon is presented, focusing on 
acceleration in relativistic shocks and having in mind a model where
the effective parameter driving LSV would vary like the square of the
momentum scale (Quadratically Deformed Relativistic Kinematics, QDRK).
Implications for ultra-high energy (UHE) 
neutrino production are equally discussed.
\end{abstract}

\section*{Introduction}

Understanding the origin of the highest-energy cosmic-ray events (in the region
$E~\approx ~10^{20}~eV$) remains a major scientific challenge, both from the 
point of view of acceleration and from that of propagation.  
Models of ultra-high energy (UHE) 
cosmic-ray acceleration seem to be confronted to 
problems 
in explaining the highest-energy cosmic-ray events, mainly due to the forces 
(magnetic fields) and source sizes required.
Synchrotron radiation (up to $\approx TeV$ photons from $\approx 10^{19}~eV$
protons) plays a crucial role,  
as energy losses prevent protons from reaching higher energies.
The theory of synchrotron radiation by electrons 
can be found in \cite{guin65} and \cite{blum70} . 
Rescaling parameters to replace the electron by a proton, we get a Larmor frequency
$\nu_{Lp}~ =~e~B~(2~\pi~m_p~c)^{-1}$ where $e$ is the electron charge, 
$B$  the magnetic field, $c$ the
speed of light and $m_p$
the proton mass. The characteristic frequency of synchrotron 
radiation is then: $\nu_{SRp}~=~3/2 ~ \nu_{Lp} ~E^2~(m_p c^2)^{-2}$ where $E$
is the proton energy.
The basic 
mechanism can be described as follows: the proton is accelerated by the
absorption of virtual photons but, for kinematical 
reasons, it becomes less and less stable under the action of the magnetic field. This 
is due to the dominance of longitudinal energy in the high-energy relation
between $E$ and $p$ (momentum):  
\begin{equation}
E~\simeq ~p~c~+~m_p^2~c^3~(2p)^{-1}
\end{equation}
As energy increases, the mass term
$m_p^2~c^3~(2p)^{-1}$ decreases linearly and it costs less and less 
energy to split the energy-momentum quadrivector of the accelerated proton, 
allowing it to decay into
a slower proton plus a longitudinally emitted photon. As pointed out in \cite{gon00} , 
Lorentz symmetry violation at the Planck scale 
can dramatically change the kinematical balance 
for this phenomenon 
and prevent synchrotron radiation emission by UHE protons in 
astrophysical sources.

\section*{Deformed relativistic kinematics (DRK)}

An example of LSV (Lorentz symmetry violation) pattern is given by 
QDRK (quadratically deformed relativistic kinematics, \cite{gon97a}). 
DRK is formulated in the vacuum rest 
frame (VRF), 
an "absolute local rest frame" expected to be close to that suggested 
by cosmic microwave background radiation. There,
the QDRK relation 
between energy and momentum in the region 
$h (2\pi~a)^{-1}~\gg~p~\gg ~mc$ where $h$ is the Planck constant
and $a$ the fundamental length, is given by \cite{gon97b}:
\begin{equation}
E~\simeq ~ p~c~+~m^2~c^3~(2~p)^{-1}~
-~p~c~\alpha ~(k~a)^2/2~~~~~
\end{equation} 
$\alpha $ being a model-dependent constant, in the range $0.1~-~0.01$ for
full-strength violation of Lorentz symmetry at the fundamental length scale.
LSV becomes phenomenologically significant above $E_{trans}~
\approx ~\pi ^{-1/2}~ h^{1/2}~(2~\alpha )^{-1/4}~a^{-1/2}~m^{1/2}~c^{3/2}$,
when the very small deformation $\Delta ~E~=~-~p~c~\alpha ~(k~a)^2/2$ 
dominates over
the mass term $m^2~c^3~(2~p)^{-1}$ and modifies all
kinematical balances. 
Such a threshold for new 
physics corresponds to
$~\approx~10^{20}~eV$
for $m$ = proton mass and
$\alpha ~a^2~\approx ~10^{-72}~cm^2$ (f.i. $\alpha
~\approx ~10^{-6}$ and $a$ = Planck
length), and to $E~\approx ~10^{18}~eV$ for
$m$ = pion mass and $\alpha ~a^2~\approx ~10^{-67}~cm^2$
(f.i. $\alpha ~\approx ~0.1$ and $a$ = Planck length).   

Assuming $c$ and $\alpha $ to be universal, and the earth to move slowly 
with respect to the VRF,
unstable particles with at
least two stable particles in the final states
of all their decay channels become stable at very
high energy. Above $E_{trans}$, the lifetimes of all
unstable particles (e.g. the $\pi ^0$ in
cascades) become much longer than predicted
by relativistic kinematics \cite{gon97a} and are expected to produce  
significant signatures \cite{gon97b} .
Then, for instance,
the neutron or even the $\Delta ^{++}$ can be suitable candidates for the
primaries of the highest-energy observed cosmic-ray
events. If $c$ and $\alpha $ are not exactly universal,
many different scenarios are possible concerning the stability of
ultra-high energy particles
\cite{gon97c} .
Similarly, the allowed final-state
phase space of two-body collisions is strongly
reduced at very high energy, leading
\cite{gon97d} to a
sharp fall of partial and total cross-sections
for incoming cosmic-ray energies above
$E_{lim} ~\approx ~(2~\pi )^{-2/3}~(E_T~a^{-2}~ \alpha ^{-1}~h^2~c^2)^{1/3}$,
where $E_T$ is the target energy.

\subsection*{DRK prevents synchrotron radiation at ultra-high energy}

Assuming $\alpha $ to be positive (otherwise LSV would lead to spontaneous 
decays at ultra-high energy \cite{gon97c} ) ,
it costs more and more energy to split the incoming longitudinal
momentum as energy increases above $E_{trans}$.

A rough example of the effect of DRK can be given as follows. 
If relativistic kinematics applies, a UHE proton with energy-momentum 
$\simeq ~[~p~c~+~m^2~c^3~(2p)^{-1}~ ,~p~]$ can 
emit in the longitudinal direction a photon with energy 
$(\epsilon ~,~ \epsilon ~c^{-1})$ if, for instance, 
it absorbs an energy $\delta ~E~\simeq ~ m^2~c^2~p^{-2}~\epsilon /2$ .
This expression for $\delta ~E$ falls quadratically with the incoming energy.  
With QDRK and above $E_{trans}$ , we get instead $\delta ~E~\simeq ~
3~\alpha ~\epsilon ~(k~a)^2/2$~, quadratically rising with proton energy.
At high enough energy, the proton can no longer emit synchrotron radiation
apart from (comparatively) very small energy losses. 
We therefore expect protons to be accelerated to higher energies in the 
presence of Lorentz symmetry violation. 

\section*{UHE pion photoproduction}

It has been suggested \cite{wax99} that, in reverse shocks,
protons accelerated to energies $\sim ~10^{20}~eV$ collide with
ambient photons producing, in the kinematical region where the center of
mass energy corresponds to the $\Delta ^+$ resonance, charged pions with about
20$\%$ of the initial proton energy. These pions subsequently decay into muons
and muon neutrinos ($\pi^+~\rightarrow~\mu^+~+~\nu_\mu $).
Muons decay later into electrons, electron neutrinos
and muon antineutrinos ($\mu^+~\rightarrow ~e^+~+~\nu_e ~+~\overline\nu_\mu $).
The authors infer from this analysis
that it may be possible, in forthcoming very large volume experiments
on earth, to observe UHE neutrinos from 
gamma-ray bursts.

It should be noticed, however, that, beacuse of Lorentz dilation,
the lifetime for the first decay at pion energy $\sim 2.10^{19}~eV$ is
$\sim 3000~ s$ . Furtermore,
the muon lifetime at $\sim 10^{19}~eV$ is $\sim 10^5~s$ .
These time scales are 
longer than those of gamma-ray bursts. But the situation,
for pion photoproduction and decay, may be considerably 
worsened by LSV \cite{gon00} . For
$\alpha ~a^2~>~10^{-72}~cm^2$ in the above proposed QDRK pattern, 
the lifetime of a $\sim ~10^{20}~eV$
$\Delta ^+$ is modified
by QDRK and gets much longer. The allowed final phase state for the reaction:
$p ~+ ~\gamma ~\rightarrow ~p ~+~X$ where $X$ is any particle or set of particles,
becomes more and more restricted (see \cite{gon97d}) and 
allows for a less and less significant energy loss 
by the incoming UHE proton.
Therefore, the 
QDRK scenario tends to preserve UHE protons and to strongly inhibit pion 
photoproduction by such protons (in agreement with data above the GZK cutoff),
as well as subsequent decays. Furthermore, as previously stated, the 
pion lifetime becomes much longer than expected from special relativity. 

\section*{Neutrino physics}

Therefore \cite{gon00} , with QDRK, 
UHE neutrino production from protons accelerated to UHE in relativistic shocks
would be inhibited by the phenomenon 
we just described. 
If $\alpha ~a^2~>~10^{-70}~cm^2$ , the $\Delta ^+$ resonance
becomes stable at $E~\sim ~10^{20}~eV$ (up to a very small tail)
with respect to the
$n~e^+~\nu _{\mu }~\overline\nu_ {\mu}~\nu _e$ decay channel,
up to a small resonance tail to be discussed in terms of the reaction
$p ~+ ~\gamma ~\rightarrow ~p ~+~X$ .

As QDRK makes the velocity $v~=~dE/dp$ of massless particles 
energy-dependent 
\cite{gon97a} , 
the arrival time on
earth of photons or neutrinos produced in a single burst
is expected to depend on the particle
energy.
But, contrary to the claim presented in \cite{el99} , it follows from the
above considerations that observing
UHE neutrinos from gamma-ray bursts for QDRK with $\alpha ~a^2~>~10^{-72}~cm^2$ 
would be much more difficult (if at all possible) than 
naively expected from the Waxman-Bahcall model. 

\section*{Concluding remarks}

Lorentz symmetry violation is not an "exotic" issue. From a strictly scientific
point of view, it is much more arbitrary to assume that special relativity 
holds up to Planck scale than to accept the simple idea presented in this 
paper, i.e. 
that we should expect it to be 
violated at the fundamental length scale. No consistent quantum field
theoretical scheme incorporating gravitation 
can prevent Lorentz symmetry violation
at the Planck scale, whereas deformed Lorentz symmetry appears 
to be consistent with all basic principles of quantum field theory.
As a consequence, the implications
of deformed Lorentz symmetry cannot be ignored when considering acceleration
to ultra-high energy in relativistic shocks.  

QDRK is a rather soft LSV scenario, as it implies a $10^{-16}$ factor between 
LSV at the $10^{20}~eV$ scale and LSV at the Planck scale. Even so, it leads to
very important effects ont the basic mechanisms of UHE cosmic-ray acceleration
and propagation, 
and can completely change all existing phenomenological predictions. The fact 
that a very small LSV can lead to observable experimental effects makes,
potentially, the 
study of UHE cosmic rays a powerful and unique
microscope directly focused on Planck-scale 
physics. The main drawback, by now, is that the LSV parameters remain strongly 
model-dependent and require a more precise dynamical scenario. However, 
phenomenological studies excluding models and values of parameters are already 
possible and can be particularly fruitful in connexion
with the forthcoming UHE cosmic-ray 
experiments (AUGER, satellite experiments... \cite{NAG}).

\end{document}